\documentclass[aps,pra,preprintnumbers,showpacs,tightenlines]{revtex4}
\usepackage{amssymb}
\usepackage{amsmath}
\usepackage{graphicx}
\usepackage{epsfig}
\usepackage{subfigure}
\usepackage{amsfonts}
\usepackage{CJK}

\begin{document}

\title{Realizing an $n$-target-qubit controlled phase gate in cavity QED:
An approach without classical pulses}

\author{Qi-Ping Su $^{1}$, Man Liu$^{1}$, and Chui-Ping Yang $^{1,2}$}
\email{yangcp@hznu.edu.cn}
\address{$^1$Department of Physics, Hangzhou Normal University,
Hangzhou, Zhejiang 310036, China}

\address{$^2$State Key Laboratory of Precision Spectroscopy, Department of Physics,
East China Normal University, Shanghai 200062, China}

\date{\today}

\begin{abstract}
We propose a way to realize a multiqubit controlled phase gate
with one qubit simultaneously controlling $n$ target qubits using
atoms in cavity QED. In this proposal, there is no need of using
classical pulses during the entire gate operation. The gate operation
time scales as $\sqrt{n}$ only and thus the gate can be performed
faster when compared with sending atoms through the cavity one at a
time. In addition, only three steps of operations are required for realizing this
$n$-target-qubit controlled phase gate. This proposal is quite
general, which can be applied to other physical systems such as
various superconducting qubits coupled to a resonator, NV centers
coupled to a microsphere cavity or quantum dots in cavity QED.
\end{abstract}

\maketitle
\date{\today}

\begin{center}
\textbf{I. INTRODUCTION}
\end{center}

Quantum computing has attracted much
attention since quantum computers can in principle solve
computational problems much more efficiently than classical
computers or process some computational tasks that are intractable
with their classical counterparts [1-3]. In the past decade,
various physical systems have been considered for building up
quantum-information processors. Among them, the cavity QED with
neutral atoms is a very promising approach for quantum information
processing, because a cavity can act as a quantum bus to couple
atoms and information can be stored in certain atomic energy
levels with long coherence time.

It is known that a quantum computation network can be constructed using
one-qubit and two-qubit logic gates [4,5]. Based on cavity QED technique,
many theoretical methods have been proposed for implementing a two-qubit
controlled-phase or controlled-NOT gate with atoms [6-12]. Moreover, a
two-qubit quantum controlled phase gate between a cavity mode and an atom
has been experimentally demonstrated [13].

Research on quantum computing has recently moved toward the
physical realization of multiqubit controlled quantum gates, which
are useful in quantum information processing. In principle, any
multiqubit gate can be decomposed into two-qubit gates and
one-qubit gates. However, when using the conventional
gate-decomposition protocols to build up a multiqubit controlled
gate [5], the procedure usually becomes complicated as the number
$n$ of qubits increases and the single-qubit and two-qubit gates
required for the gate implementation heavily depends on the number
$n$ of qubits (especially, for a large $n$). Therefore, it is
important to find a more efficient way to realize multiqubit
controlled gates.

\begin{figure}[tbp]
\begin{center}
\includegraphics[bb=143 345 509 574, width=7.5 cm, clip]{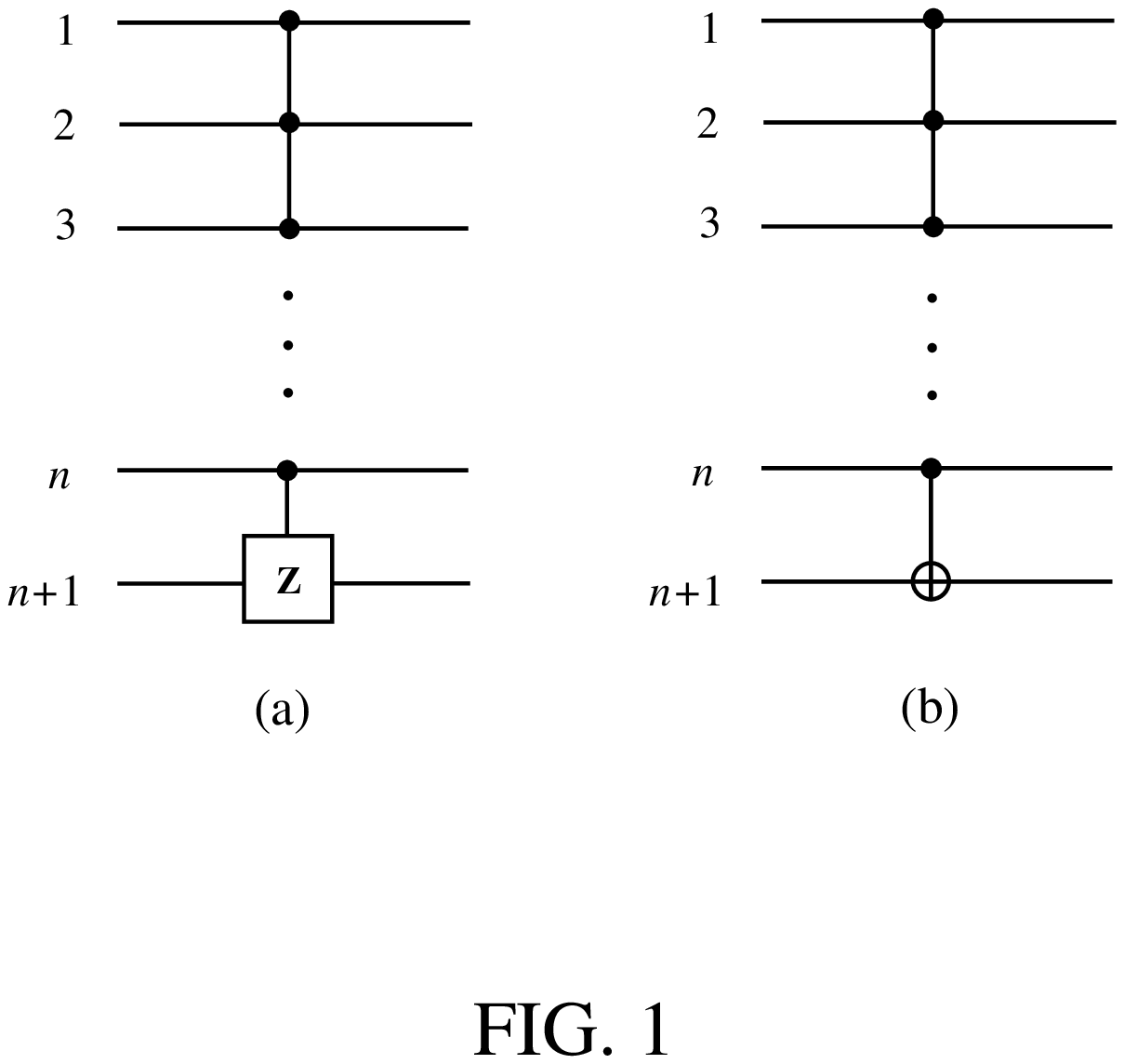} %
\vspace*{-0.08in}
\end{center}
\caption{(a) Circuit of a controlled-phase gate with $n$ control
qubits ($1,2,...,n$) acting on a target qubit $n+1$. The symbol
$Z$ represents a controlled-phase shift. If the $n$ control qubits
(linked with the filled circles) are \textit{all} in the state
$\left| 1\right\rangle $, then the state $\left| 1\right\rangle $
of the target qubit located at the bottom is phase-shifted by
$\pi$ (i.e., $\left| 1\right\rangle \rightarrow e^{i\pi}\left|
1\right\rangle=-\left| 1\right\rangle $); otherwise nothing
happens to the states of the target qubit. (b) Circuit of a
controlled-NOT gate with $n$ control qubits ($1,2,...,n$) acting
on a target qubit $n+1$. The symbol $\oplus $ represents a
controlled-NOT gate (with $n$ control qubits on the filled
circles). If the $n$ control qubits are \textit{all} in the state
$\left| 1\right\rangle $, then the state of the target qubit at
the bottom is bit-flipped (i.e., $\left| 1\right\rangle
\rightarrow \left| 0\right\rangle $ and $\left| 0\right\rangle
\rightarrow \left| 1\right\rangle $).} \label{fig:1}
\end{figure}

During the past few years, many schemes have been proposed for
implementing multiqubit controlled gates in different physical
systems, e.g., atoms in cavity QED [14,15], trapped ions [16],
atomic ensembles [17], superconducting qubits coupled to a cavity
or resonator [18,19], and nitrogen-vacancy (NV) centers [20]. The
proposals [14-20] are mainly for implementing a multiqubit
controlled-phase or controlled-NOT gate with
\textit{multiple-control} qubits acting on \textit{one target}
qubit (Fig.~1). This type of multiqubit controlled gates is of
significance in quantum information processing such as quantum
algorithms (e.g., [2,21]) and quantum error-correction protocols
[22].

In this work, we focus on another type of multiqubit controlled
gates, i.e, a multiqubit phase or CNOT gate with one qubit
simultaneously controlling $n$ target qubits (Fig.~2). This type
of controlled gates with $n$ target qubits is useful in quantum
information processing. For instances, they have applications in
error correction~[23], quantum algorithms (e.g., the Discrete
Cosine Transform~[24]), and quantum cloning~[25]. In addition,
they can be used to prepare Greenberger-Horne-Zeilinger (GHZ)
states [26]. It should be mentioned here that the gate in Fig.~1
can be used to prepare states which are locally-equivalent to
W-class states [27], but can not be applied to create GHZ states.
As is well known, the GHZ states and the W-class states can not be
interchanged to each other, and both of them play an important
role in quantum information processing and communication.

\begin{figure}[tbp]
\begin{center}
\includegraphics[bb=129 179 435 620, width=7.5 cm, clip]{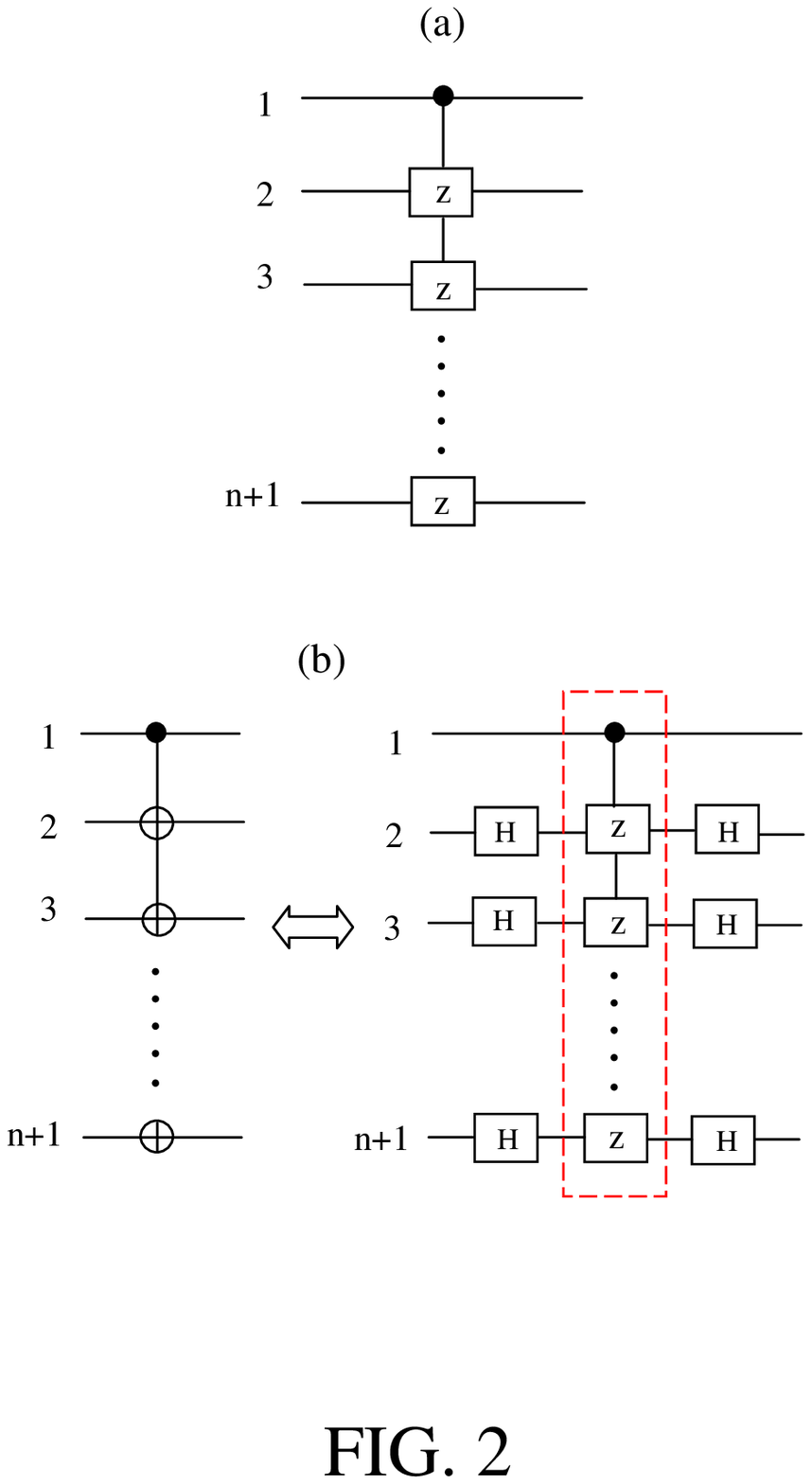}
\vspace*{-0.08in}
\end{center}
\caption{(Color online) (a) Circuit of a phase gate with qubit 1
simultaneously controlling $n$ target qubits (2,~3,~...,~$n+1$).
This $n$-target-qubit controlled phase gate is equivalent to $n$
two-qubit control phase (CP) gates each having a shared control
qubit (qubit 1) but a different target qubit (qubit
2,~3,~...,~or~$n+1$). Here, Z represents a controlled-phase
flip on each target qubit.~Namely, if the control qubit 1 is in the state $%
\left| 1\right\rangle $, then the state $\left| 1\right\rangle $
at each Z is phase-flipped as $\left| 1\right\rangle $
$\rightarrow -\left| 1\right\rangle $, while the state $\left|
0\right\rangle $ remains unchanged.~(b) Relationship between an
$n$-target-qubit controlled-NOT gate and the $n$-target-qubit
controlled phase gate. The circuit on the left side of (b) is
equivalent to the circuit on the right side of (b). For the
circuit on the left side, the symbol $\oplus $ represents a CNOT
gate on each target qubit.~If the control qubit 1 is in the state
$\left| 1\right\rangle $, then
the state at $\oplus $ is bit flipped as $\left| 1\right\rangle $ $%
\rightarrow \left| 0\right\rangle $ and $\left| 0\right\rangle $ $%
\rightarrow \left| 1\right\rangle $.~However,~when the control
qubit 1 is in the state $\left| 0\right\rangle $,~the state at
$\oplus $ remains unchanged. On the other hand, for the circuit on
the right side, the part enclosed in the (red) dashed-line box
represents the $n$-target-qubit controlled phase gate shown in
(a). The element containing H corresponds to a Hadamard
transformation described by $\left| 0\right\rangle \rightarrow \left( 1/%
\protect\sqrt{2}\right) \left( \left| 0\right\rangle +\left|
1\right\rangle
\right) $, and $\left| 1\right\rangle \rightarrow \left( 1/\protect\sqrt{2}%
\right) \left( \left| 0\right\rangle -\left| 1\right\rangle
\right) .$} \label{fig:2}
\end{figure}

For simplicity, we denote this multiqubit phase gate with one
qubit simultaneously controlling $n$ target qubits [see Fig.~2(a)]
as {\it an $n$-target-qubit controlled phase gate}. It can be seen
from Fig.~2(a) that the $n$-target-qubit controlled phase gate
consists of $n$ two-qubit controlled-phase (CP) gates. Each
two-qubit CP gate involved in this multiqubit gate has a
\textit{shared} control qubit (labelled by $1$) but a different
target qubit (labelled by $2,3,...,$ or $n+1$). For two qubits,
there are a total of four computational basis states $\left|
00\right\rangle ,$ $\left| 01\right\rangle ,$ $\left|
10\right\rangle ,$ and $\left|
11\right\rangle .$ The two-qubit CP gate acting on qubit $1$ and qubit $j$ ($%
j=2,3,...,n+1$) is defined as $\left| 0\right\rangle _1\left| 0\right\rangle
_j\rightarrow \left| 0\right\rangle _1\left| 0\right\rangle _j,\;\left|
0\right\rangle _1\left| 1\right\rangle _j\rightarrow \left| 0\right\rangle
_1\left| 1\right\rangle _j,\;\left| 1\right\rangle _1\left| 0\right\rangle
_j\rightarrow \left| 1\right\rangle _1\left| 0\right\rangle _j,\;$and $%
\left| 1\right\rangle _1\left| 1\right\rangle _j\rightarrow
-\left| 1\right\rangle _1\left| 1\right\rangle _j,$ which implies
that if and only if the control qubit $1$ is in the state $\left|
1\right\rangle $, a phase flip happens to the state $\left|
1\right\rangle $ of the target qubit $j$, but nothing happens
otherwise. According to the definition of a two-qubit CP gate
here, it is easy to see that this $n$-target-qubit controlled
phase gate with one qubit $1$ simultaneously controlling $n$
target qubits ($2,3,...,n+1$) is described by the following
unitary operator
\begin{equation}
U_p=\prod_{j=2}^{n+1}\left( I_j-2\left| 1\right\rangle _1\left|
1\right\rangle _j\left\langle 1\right| _1\left\langle 1\right| _j\right) ,
\end{equation}
where the subscript $1$ represents the control qubit $1$, while $j$
represents the target qubit $j$; and $I_j$ is the identity operator for the
qubit pair ($1,j$), which is given by $I_j=\sum_{\mathrm{rs}}\left| \mathrm{r%
}\right\rangle _1\left| \mathrm{s}\right\rangle _j\left\langle \mathrm{r}%
\right| _1\left\langle \mathrm{s}\right| _j$, with
$\mathrm{r,s}\in \{0,1\}.$ From Eq.~(1), it can be seen that the
operator $U_p$ induces a phase flip (from the $+$ sign to the $-$
sign) to the logical state $\left| 1\right\rangle $ of each target
qubit when the control qubit $1$ is initially in the state $\left|
1\right\rangle $, and nothing happens otherwise.

In this paper, we will present a way for implementing the
$n$-target-qubit controlled phase gate with ($n+1$) atoms in
cavity QED. Here, the ($n+1$) atoms are one control atom acting as
a control qubit and $n$ target atoms each playing a role of a
target qubit. This proposal has the following features: (i) there
is no need of using classical pulses during the entire operation;
(ii) The gate operation time scales as $\sqrt{n}$ only and thus the
gate can be performed faster when compared with sending atoms through
the cavity one by one; (iii) the $n$ two-qubit CP gates involved can
be simultaneously performed; and (v) the gate implementation
requires only three steps of operations. This proposal is quite general, which can be
applied to other physical systems such as various superconducting
qubits coupled to a resonator, nitrogen-vacancy (NV) centers
coupled to a microsphere cavity or quantum dots in cavity QED.

Note that an $n$-target-qubit CNOT gate, shown in Fig.~2(b), can
also be achieved using the present proposal. This is because the
n-target-qubit CNOT gate is equivalent to the $n$-target-qubit
controlled phase gate discussed above, plus two Hadamard gates on
each target qubit [Fig.~2(b)].

This paper is organized as follows. In Sec.~II, we briefly review the basic
theory of atom-cavity resonant interaction and atom-cavity off-resonant
interaction. In Sec.~III, we show how to realize an $n$-target-qubit
controlled phase gate using atoms in cavity QED. In Sec.~IV, we study
fidelity of the gate operation. A brief discussion and the summary are given
in Sec.~V.

\begin{center}
\textbf{II. BASIC THEORY}
\end{center}

The gate implementation requires two types
of atom-cavity interaction, which are described as follows.

\textit{A. Atom-cavity resonant interaction.} Consider a two-level
atom, say, atom $1$, with a ground level $\left| 0\right\rangle $
and an excited level $\left| 1\right\rangle .$ Assume that the
cavity mode is resonant with the $\left| 0\right\rangle
\leftrightarrow \left| 1\right\rangle $ transition of the atom.
The interaction Hamiltonian in the interaction picture can be
written as
\begin{equation}
H=\hbar g_r(a^{+}\sigma _{01}^{-}+ H.c.),
\end{equation}
where $a^{+}$ and $a$ are the photon creation and annihilation operators of
the cavity mode, $g_r$ is the \textit{resonant} coupling constant between
the cavity mode and the $\left| 0\right\rangle \leftrightarrow \left|
1\right\rangle $ transition of the atom, and $\sigma _{01}^{-}=\left|
0\right\rangle \left\langle 1\right| $. It is easy to find that the time
evolution of the states $\left| 1\right\rangle \left| 0\right\rangle _c$ and
$\left| 0\right\rangle \left| 1\right\rangle _c$ of the atom and the cavity
mode, governed by the Hamiltonian~(2), is described by
\begin{eqnarray}
\left| 1\right\rangle \left| 0\right\rangle _c &\rightarrow &-i\sin \left(
g_rt\right) \left| 0\right\rangle \left| 1\right\rangle _c+\cos (g_rt)\left|
1\right\rangle \left| 0\right\rangle _c,  \nonumber \\
\left| 0\right\rangle \left| 1\right\rangle _c &\rightarrow &\cos \left(
g_rt\right) \left| 0\right\rangle \left| 1\right\rangle _c-i\sin \left(
g_rt\right) \left| 1\right\rangle \left| 0\right\rangle _c,
\end{eqnarray}
while the state $\left| 0\right\rangle \left| 0\right\rangle _c$ remains
unchanged. Here and below, the $\left| 0\right\rangle _c$ and $\left|
1\right\rangle _c$ are the vacuum state and the single-photon state of the
cavity mode, respectively.

\textit{B. Atom-cavity off-resonant interaction.} Consider $n$
atoms ($2,3,...,n+1$) each having three levels $\left|
0\right\rangle ,$ $\left| 1\right\rangle ,$ and $\left|
2\right\rangle $ (Fig.~3). Suppose that the cavity mode is coupled
to the $\left| 1\right\rangle \leftrightarrow \left|
2\right\rangle $ transition of each atom but highly detuned
(decoupled) from the transition between any other two levels
(Fig.~3). In the interaction picture, the interaction Hamiltonian
of the whole system is given by

\begin{equation}
H=\hbar \sum_{k=2}^{n+1}g(e^{-i\Delta _ct}a^{+}\sigma _{12,k}^{-}+ H.c.),
\end{equation}
where $\Delta _c=\omega _{21}-\omega _c$ is the detuning between the cavity
mode frequency $\omega _c$ and the $\left| 1\right\rangle \leftrightarrow
\left| 2\right\rangle $ transition frequency $\omega _{21}$ of the atoms, $g$
is the coupling constant between the cavity mode and the $\left|
1\right\rangle \leftrightarrow \left| 2\right\rangle $ transition, and $%
\sigma _{12,k}^{-}=\left| 1\right\rangle _k\left\langle 2\right| .$

\begin{figure}[tbp]
\begin{center}
\includegraphics[bb=236 336 404 556, width=3.5 cm, clip]{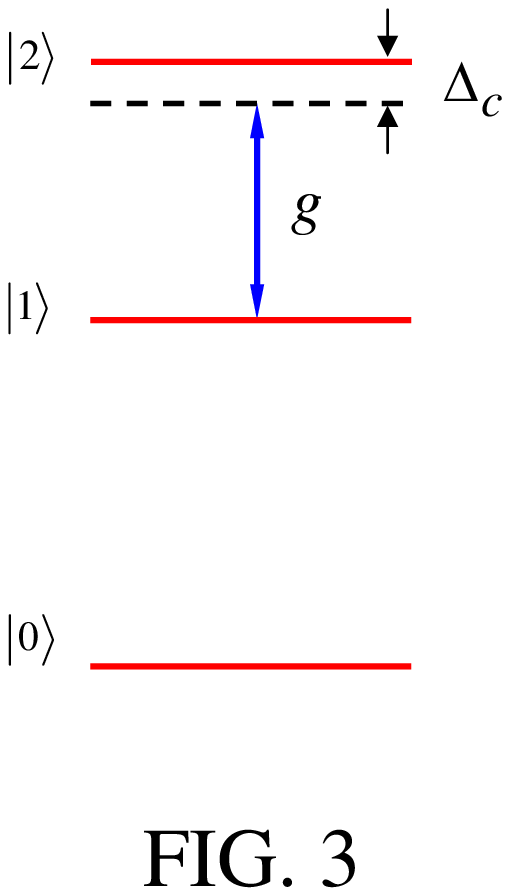} %
\vspace*{-0.08in}
\end{center}
\caption{(Color online) Atom-cavity off-resonant interaction. The
cavity mode is off-resonant with the $\left| 1\right\rangle
\leftrightarrow \left|
2\right\rangle $ transition with a detuning $\Delta _c=\omega_{21}-\omega_c$%
. Here, $\omega_c$ is the cavity mode frequency while $\omega_{21}$ is the $%
\left| 1\right\rangle \leftrightarrow \left| 2\right\rangle $
transition frequency of the atom. In addition, $g$ is the
off-resonant coupling constant between the cavity mode with the
$\left| 1\right\rangle \leftrightarrow \left| 2\right\rangle $
transition. To have the quantum information of a qubit to be
stored in the two lowest levels for a long time, the atoms can be
chosen for which the transition between the two lowest levels is
forbidden due to the selection rule, or the two lowest levels are
chosen to be hyperfine levels of an atom.} \label{fig:3}
\end{figure}

For the case of $\Delta _c\gg \sqrt{n}g$ (i.e., the cavity mode is
off-resonant with the $\left| 1\right\rangle \leftrightarrow
\left| 2\right\rangle $ transition of each atom), there is no
energy exchange between the atoms and the cavity mode. However,
the coupling between the atoms and the cavity mode may induce a phase to
the atomic states. This kind of coupling, which can induce
a phase but does not cause energy exchange, is often called ``dispersive coupling"
in quantum optics, or it is said that the cavity mode is dispersively coupled
to the atoms. In this case, based on the Hamiltonian (4), one can
obtain the following effective Hamiltonian [6,28,29]
\begin{eqnarray}
H_{\mathrm{eff}} &=&-\hbar \sum_{k=2}^{n+1}\frac{g^2}{\Delta _c}\left(
a^{+}a\sigma _{11,k}-aa^{+}\sigma _{22,k}\right)  \nonumber \\
&&\ \ +\hbar \sum_{k\neq k^{\prime }=2}^{n+1}\frac{g^2}{\Delta _c}\left(
\sigma _{12,k}^{+}\sigma _{12,k^{\prime }}^{-}+\sigma _{12,k}^{-}\sigma
_{12,k^{\prime }}^{+}\right) ,
\end{eqnarray}
where the two terms in the first line represent the photon-number-dependent
Stark shifts, while the two terms in the second line describe the ``dipole''
coupling between the two atoms ($k,k^{\prime }$) mediated by the cavity
mode. When the level $\left| 2\right\rangle $ of each atom is not occupied,
the Hamiltonian (5) reduces
\begin{equation}
H_{\mathrm{eff}}=-\hbar \sum_{k=2}^{n+1}\frac{g^2}{\Delta _c}a^{+}a\sigma
_{11,k}.
\end{equation}

The time-evolution operator for the Hamiltonian (6) is
\begin{equation}
U(t)=\otimes _{k=2}^{n+1}U_{kc}\left( t\right) .
\end{equation}
Here, $U_{kc}\left( t\right) $ is the time-evolution operator acting on the
cavity mode and the atom $k$ ($k=2,3,...,n+1$), which is given by
\begin{equation}
U_{kc}\left( t\right) =\exp [i\left( g^2/\Delta _c\right) a^{+}a\sigma
_{11,k}t].
\end{equation}
One can easily find that the operator $U_{kc}\left( t\right) $ results in
the following state transformation
\begin{eqnarray}
\left| 0\right\rangle _k\left| 0\right\rangle _c &\rightarrow &\left|
0\right\rangle _k\left| 0\right\rangle _c,  \nonumber \\
\left| 1\right\rangle _k\left| 0\right\rangle _c &\rightarrow &\left|
1\right\rangle _k\left| 0\right\rangle _c,  \nonumber \\
\left| 0\right\rangle _k\left| 1\right\rangle _c &\rightarrow &\left|
0\right\rangle _k\left| 1\right\rangle _c,  \nonumber \\
\left| 1\right\rangle _k\left| 1\right\rangle _c &\rightarrow &e^{i\phi
_k\left( t\right) }\left| 1\right\rangle _k\left| 1\right\rangle _c,
\end{eqnarray}
where $\phi _k\left( t\right) =g^2t/\Delta _c.$ This result~(9) demonstrates
that a phase shift $\phi _k\left( t\right) $ is induced to the state $\left|
1\right\rangle $ of the atom $k$ in the case when the cavity mode is in the
single-photon state $\left| 1\right\rangle _c.$ For $t=\pi \Delta _c/g^2,$
we have $\phi _k\left( t\right) =\pi ,$ i.e., $\left| 1\right\rangle
_k\left| 1\right\rangle _c\rightarrow -\left| 1\right\rangle _k\left|
1\right\rangle _c,$ which implies that a phase flip is induced to the state $%
\left| 1\right\rangle $ of the atom $k$ by the cavity photon. In addition,
the first two lines of Eq.~(9) show that the states $\left| 0\right\rangle $
and $\left| 1\right\rangle $ of atom $k$ remain unchanged when the cavity
mode is in the vacuum state $\left| 0\right\rangle _c,$ and the third line
of Eq.~(9) shows that the state $\left| 0\right\rangle $ of atom $k$ remains
unchanged when the cavity mode is in the single-photon state $\left|
1\right\rangle _c.$

The operator $U(t)$ is a product of the operators $U_{2c}\left( t\right) ,$ $%
U_{3c}\left( t\right) ,...,$ and $U_{(n+1)c}\left( t\right) ,$ which can be
seen from Eq.~(7)$.$ Based on Eq.~(7), Eq.~(8) and the result~(9), one can
easily find:

(i) The states $\left| 0\right\rangle $ and $\left| 1\right\rangle $ of each
of atoms ($2,3,...,n+1$) remain unchanged when the cavity mode is in the
vacuum state $\left| 0\right\rangle _c;$

(ii) The state $\left| 0\right\rangle $ of each of atoms ($2,3,...,n+1$)
remains unchanged when the cavity mode is in the single-photon state $\left|
1\right\rangle _c;$

(iii) For $t=\pi \Delta _c/g^2,$ a phase flip happens to the state $\left|
1\right\rangle $ of each of atoms ($2,3,...,n+1$) simultaneously, in the
case when the cavity mode is in the single-photon state $\left|
1\right\rangle _c.$ To see this, consider the state $\left| \varphi
\right\rangle =\left| 1\right\rangle _2\left| 1\right\rangle _3\left|
1\right\rangle _4\left| 1\right\rangle _c$ for three atoms\ ($2,3,4$) and
the cavity mode. One can easily verify that by applying the operator $%
U(t)=U_{2c}\left( t\right) $ $U_{3c}\left( t\right) U_{4c}\left( t\right) $
to the state $\left| \varphi \right\rangle ,$ the state $\left| \varphi
\right\rangle $ becomes $(-\left| 1\right\rangle _2)(-\left| 1\right\rangle
_3)(-\left| 1\right\rangle _4)\left| 1\right\rangle _c$ for $t=\pi \Delta
_c/g^2,$ which shows that a phase flip is induced to the state $\left|
1\right\rangle $ of each of three atoms ($2,3,4$) at the same time, by the
cavity photon.

The results (i-iii) given here will be applied to the gate implementation
discussed in next section.

\begin{center}
\textbf{III. IMPLEMENTATION OF AN N-TARGET-QUBIT CONTROLLED PHASE GATE}
\end{center}

To realize the gate, we will employ a two-level atom $1$
and $n$ identical
three-level atoms ($2,3,...,n+1$). The three levels of each of atoms ($%
2,3,...,n+1$) are shown in Fig.~3. For each atom, the two lowest levels $%
\left| 0\right\rangle $ and $\left| 1\right\rangle $ represent the two
logical states of a qubit. In the following, atom $1$ acts as a \textit{%
control} while each one of the atoms ($2,3,...,n+1$) plays a \textit{target}
role.

We suppose that during the following gate operation, (i) the cavity mode is
resonant with the $\left| 0\right\rangle \rightarrow $ $\left|
1\right\rangle $ transition of atom $1;$ and (ii) the cavity mode is
off-resonance with the $\left| 1\right\rangle \leftrightarrow \left|
2\right\rangle $ transition of atoms ($2,3,...,n+1$) but highly detuned
(decoupled) from the transition between any other two levels of atoms ($%
2,3,...,n+1$). These conditions can in principle be met by prior
adjustment of the cavity mode frequency or by appropriately
choosing atoms to have the desired level structure. Note that the
cavity mode frequency for both optical cavities and microwave
cavities can be changed in various experiments (e.g., see
[30-33]).

The cavity mode is initially in the vacuum state $\left|
0\right\rangle _c.$ The procedure for implementing the
$n$-target-qubit controlled phase gate described by Eq.~(1) is
listed below:

\textit{Step (i)}: Send atom $1$ through the cavity for an interaction time $%
t_1=\pi /(2g_r)$ by choosing the atomic velocity appropriately. After atom $%
1 $ exits the cavity, the state $\left| 0\right\rangle \left| 0\right\rangle
_c $ for atom $1$ and the cavity mode remains unchanged, while their state $%
\left| 1\right\rangle \left| 0\right\rangle _c$ changes to $-i\left|
0\right\rangle \left| 1\right\rangle _c$ as described by Eq.~(3).

\textit{Step (ii):} Send atoms ($2,3,...,n+1$) through the cavity for an
interaction time $t_2=\pi \Delta _c/g^2$ by choosing the atomic velocity
appropriately. After atoms ($2,3,...,n+1$) leave the cavity, (i) the states $%
\left| 0\right\rangle $ and $\left| 1\right\rangle $ of each atom remain
unchanged in the case when the cavity mode is in the vacuum state $\left|
0\right\rangle _c;$ (ii) the state $\left| 0\right\rangle $ of each atom
remains unchanged when the cavity mode is in the single-photon state $\left|
1\right\rangle _c;$ but (iii) the state $\left| 1\right\rangle $ of each
atom changes to $-\left| 1\right\rangle $ (i.e., a phase flip happens to the
state $\left| 1\right\rangle $ of each of atoms $2,3,...,$ and $n+1$) when
the cavity mode is in the single-photon state, as discussed in the previous
subsection B.

\textit{Step (iii):} Send atom $1$ back through the cavity for an
interaction time $t_3=3\pi /(2g_r)$. After atom $1$ leaves the cavity, the
state $\left| 0\right\rangle \left| 0\right\rangle _c$ of atom $1$ and the
cavity mode remains unchanged, while their state $\left| 0\right\rangle
\left| 1\right\rangle _c$ changes to $i\left| 1\right\rangle \left|
0\right\rangle _c.$

Note that the level $\left| 0\right\rangle $ of each of atoms ($2,3,...,n+1$%
) is not affected by the cavity mode during the operation of step (ii). This
is because the cavity mode was assumed to be highly detuned (decoupled) from
the $\left| 0\right\rangle \leftrightarrow \left| 1\right\rangle $
transition and the $\left| 0\right\rangle \leftrightarrow \left|
2\right\rangle $ transition of each of the atoms ($2,3,...,n+1$), as
mentioned before.

One can check that the $n$-target-qubit controlled phase gate,
described by Eq.~(1), was obtained with ($n+1$) atoms (i.e., the
control atom 1 and the target atoms $2,3,...,$ and $n+1$) after
the above manipulation.

Let us consider a three-qubit example in order to see better how
the multi-target-qubit controlled phase gate described by Eq.~(1)
is realized after the operations above. For three qubits, there
are a total of eight computational basis states from $\left|
000\right\rangle $ to $\left| 111\right\rangle $. Based on the
results given above for each step of operations, it can be easily
found that for three-qubit computational basis states $\left|
100\right\rangle ,\left| 101\right\rangle ,\left| 110\right\rangle $ and $%
\left| 111\right\rangle $, the time evolution of the states of the whole
system after each step of operations is given by

\begin{eqnarray}
&&\
\begin{array}{c}
\left| 1\right\rangle \left| 0\right\rangle \left| 0\right\rangle \left|
0\right\rangle _c \\
\left| 1\right\rangle \left| 0\right\rangle \left| 1\right\rangle \left|
0\right\rangle _c \\
\left| 1\right\rangle \left| 1\right\rangle \left| 0\right\rangle \left|
0\right\rangle _c \\
\left| 1\right\rangle \left| 1\right\rangle \left| 1\right\rangle \left|
0\right\rangle _c
\end{array}
\stackrel{\text{Step\thinspace (i)}}{\longrightarrow }
\begin{array}{c}
\;-i\left| 0\right\rangle \left| 0\right\rangle \left| 0\right\rangle \left|
1\right\rangle _c \\
\,\;-i\left| 0\right\rangle \left| 0\right\rangle \left| 1\right\rangle
\left| 1\right\rangle _c \\
\;\,-i\left| 0\right\rangle \left| 1\right\rangle \left| 0\right\rangle
\left| 1\right\rangle _c \\
\,\,\,-i\left| 0\right\rangle \left| 1\right\rangle \left| 1\right\rangle
\left| 1\right\rangle _c
\end{array}
\nonumber \\
&&\ \stackrel{\text{Step\thinspace (ii)}}{\longrightarrow }
\begin{array}{c}
-i\left| 0\right\rangle \left| 0\right\rangle \left| 0\right\rangle \left|
1\right\rangle _c \\
\ \ \,\;\ \ -i\left| 0\right\rangle \left| 0\right\rangle (-\left|
1\right\rangle )\left| 1\right\rangle _c \\
\ \ \ \ \ -i\left| 0\right\rangle (-\left| 1\right\rangle )\left|
0\right\rangle \left| 1\right\rangle _c \\
\ \ \ \ \ \ \ \ \ \ -i\left| 0\right\rangle (-\left| 1\right\rangle
)(-\left| 1\right\rangle )\left| 1\right\rangle _c
\end{array}
\nonumber \\
&&\stackrel{\text{Step\thinspace (iii)}}{\longrightarrow }
\begin{array}{c}
\left| 1\right\rangle \left| 0\right\rangle \left| 0\right\rangle \left|
0\right\rangle _c \\
\ \ \ \ \ \,\left| 1\right\rangle \left| 0\right\rangle (-\left|
1\right\rangle )\left| 0\right\rangle _c \\
\ \ \ \ \,\ \left| 1\right\rangle (-\left| 1\right\rangle )\left|
0\right\rangle \left| 0\right\rangle _c \\
\ \ \ \,\ \ \ \ \,\ \ \left| 1\right\rangle (-\left| 1\right\rangle
)(-\left| 1\right\rangle )\left| 0\right\rangle _c
\end{array}
.
\end{eqnarray}

Here and above, $\left| ijk\right\rangle $ is an abbreviation of the state $%
\left| i\right\rangle _1\left| j\right\rangle _2\left| k\right\rangle _3$ of
atoms ($1,2,3$) with $i,j,k\in \{0,1\}.$ The result~(10) shows that when the
control atom $1$ is initially in the state $\left| 1\right\rangle $, a phase
flip happens to the state $\left| 1\right\rangle $ of atoms ($2,3$) while
the cavity mode returns to its original vacuum state, after the above
three-step operation.

On the other hand, it is obvious that the following states of the system
\begin{equation}
\;\left| 000\right\rangle \left| 0\right\rangle _c,\left| 001\right\rangle
\left| 0\right\rangle _c,\left| 010\right\rangle \left| 0\right\rangle
_c,\left| 011\right\rangle \left| 0\right\rangle _c
\end{equation}
remain unchanged during the entire operation. This is because no
photon was emitted to the cavity during the operation of step~(i)
due to energy conservation, when atom $1$ is initially in the
state $\left| 0\right\rangle $. Hence, it can be concluded from
Eq.~(10) that a two-target-qubit controlled phase gate, i.e., a
phase gate with one qubit simultaneously controlling two target
qubits, was achieved with three atoms (i.e., the control atom $1$
and the target atoms $2$ and $3$) after the above process.

\begin{figure}[tbp]
\includegraphics[bb=306 493 528 706, width=4.0 cm, clip]{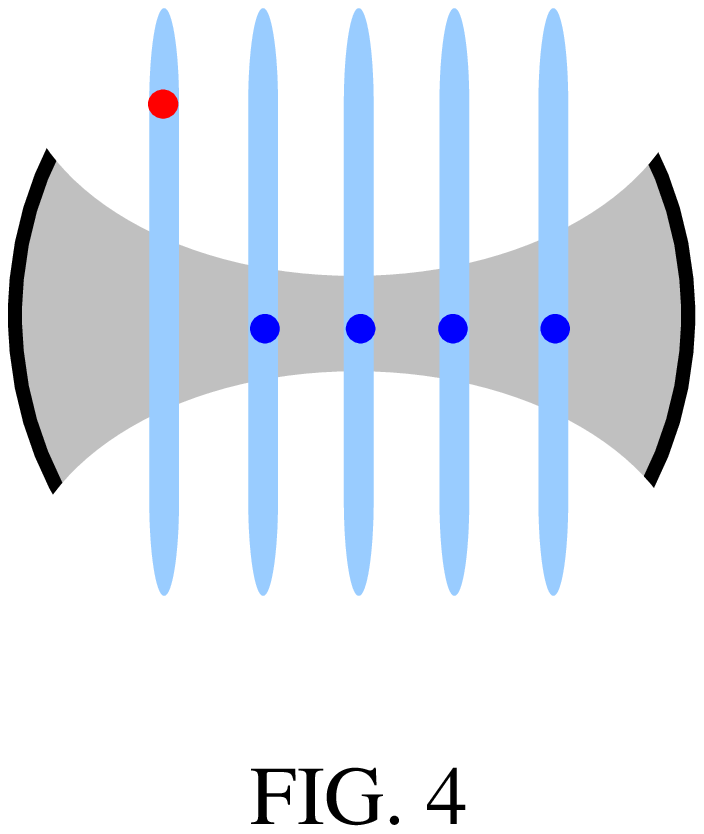} %
\vspace*{-0.08in} \caption{(Color online) Proposed setup for the
gate realization with the control atom (the red dot), the $n$
identical target atoms (the blue dots), and a cavity. For
simplicity, only five atoms are drawn here. Each atom can be either
loaded into the cavity or moved out of the cavity by one-dimensional
translating optical lattices [14,36,37].}
\label{fig:4}
\end{figure}

One can see that our gate implementation above requires sending atoms
through a cavity. We should point out that the technique by sending atoms
through a cavity to have the atoms dispersively coupled with the cavity mode
was previously proposed and has been widely used by the quantum information
community [28,34,35].

As shown above, atoms were sent through the cavity. Alternatively,
as illustrated in Fig.~4, each atom can be trapped
inside an optical lattice [14,36,37] and then can be moved into, out of,
or back into a cavity, by moving the lattice [14,36,37].

From the above descriptions, it can be seen that the total operation time $%
\tau $ is given by
\begin{equation}
\tau =2\pi /g_r+\pi \Delta _c/g^2.
\end{equation}
Because of $\Delta _c\gg g$ (set above), we can define $\Delta _c=kg$ with $%
k\gg 1.$ Thus, Eq.~(12) can be rewritten as $\tau =2\pi /g_r+k\pi
/g,$ showing that the operation time $\tau $ is independent of the
number of qubits$.$ In contrast, when employing the protocol,
which is based on sending the target atoms through the cavity one
at a time, the large detuning is $\Delta _c\gg g$ (i.e., $\Delta
_c=kg$ with $k\gg 1),$ and the total operation time is $\tau =2\pi
/g_r+kn\pi /g,$ which increases linearly with the number $n$ of
the target atoms$.$ Therefore, by using the present proposal, the
gate can be performed faster especially for a large number $n.$

\begin{center}
\textbf{IV. FIDELITY}
\end{center}

Let us now study the fidelity of the gate operations. We note that
since the resonant interaction between
the atom $1$ and the cavity is used in steps (i) and (iii), these
two steps can be completed within a very short time (e.g., by
increasing the resonant atom-cavity coupling constant $g_r$), such
that the dissipation of the atom $1$ and the cavity is negligibly
small. In this case, the dissipation of the system would appear in
the operation of step (ii) due to the use of the atom-cavity
dispersive interaction. During the operation of step (ii), the
dynamics of the lossy system, composed of the cavity mode and the
atoms ($2,3,...,n+1$), is determined by
\begin{equation}
\frac{d\rho }{dt}=-i\left[ H,\rho \right] +\kappa \mathcal{L}\left[ a\right]
+\sum_{k=2}^{n+1}\gamma _{21}\mathcal{L}\left[ \sigma _{12,k}^{-}\right]
+\sum_{k=2}^{n+1}\gamma _{20}\mathcal{L}\left[ \sigma _{02,k}^{-}\right]
+\sum_{k=2}^{n+1}\gamma _{10}\mathcal{L}\left[ \sigma _{01,k}^{-}\right] ,
\end{equation}
where $H$ is the Hamiltonian (4), $\mathcal{L}\left[ a\right] =\left( 2a\rho
a^{+}-a^{+}a\rho -\rho a^{+}a\right) ,$ $\mathcal{L}\left[ \sigma
_{ij,k}^{-}\right] =2\sigma _{ij,k}^{-}\rho \sigma _{ij,k}^{+}-\sigma
_{ij,k}^{+}\sigma _{ij,k}^{-}\rho -\rho \sigma _{ij,k}^{+}\sigma _{ij,k}^{-}$
(with $\sigma _{ij,k}^{-}=\left| i\right\rangle _k\left\langle j\right| $, $%
\sigma _{ij,k}^{+}=\left| j\right\rangle _k\left\langle i\right| $ and $%
ij\in \{12,02,01\}$), $\kappa $ is the decay rate of the cavity mode, $%
\gamma _{ji}$ is the decay rate of the level $\left| j\right\rangle $ of the
atoms ($2,3,...,n+1$) via the decay path $\left| j\right\rangle \rightarrow
\left| i\right\rangle $ (here, $ji\in \{21,20,10\}$). The fidelity of the
gate operations is given by
\begin{equation}
\mathcal{F}=\left\langle \psi _{id}\right| \widetilde{\rho }\left| \psi
_{id}\right\rangle ,
\end{equation}
where $\left| \psi _{id}\right\rangle $ is the state of the whole
system after the above three-step gate operations, in the ideal
case without considering the dissipation of the system during the
entire gate operation; and $\widetilde{\rho }$ is the final
density operator of the whole system after the gate operations are
performed in a real situation.

\begin{figure}[tbp]
\begin{center}
\includegraphics[bb=0 0 504 326, width=6.5 cm, clip]{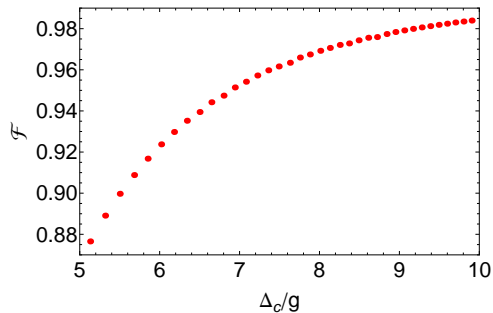} %
\vspace*{-0.08in}
\end{center}
\caption{(Color online) Fidelity of the gate operations versus the radio $%
\Delta_c/g$. The parameters used in the numerical calculation are
$\gamma _{21}^{-1}=\gamma _{20}^{-1}=\gamma _{10}^{-1}=3\times
10^{-2}$ s, $\kappa ^{-1} =3\times 10^{-2}$ s, and $g=2\pi \times
50$ KHz.} \label{fig:5}
\end{figure}

We now numerically calculate the fidelity of the gate operations.
As an example, we consider realizing a four-target-qubit
controlled phase gate, using a two-level control atom $1$ and four
identical three-level target atoms ($2,3,4,5$). The four identical
target atoms ($2,3,4,5$) are chosen as Rydberg atoms with the
principle quantum numbers 49, 50, and 51, which correspond to the
three levels $\left| 0\right\rangle ,$ $\left| 1\right\rangle ,$
and $\left| 2\right\rangle $ as depicted in Fig. 3, repectively.
For the Rydberg atoms chosen here, the energy relaxation
time $\gamma _1^{-1}$ of the level $%
\left| 1\right\rangle $ and the energy relaxation time $\gamma
_2^{-1}$ of the level $\left| 2\right\rangle $ are both $\sim
3\times 10^{-2}$ s (e.g., see [30,38,39]). Without loss of
generality, we assume that each of the five atoms is initially in
the state $\left( \left| 0\right\rangle +\left| 1\right\rangle
\right) /\sqrt{2}$ and the cavity mode is in the vacuum state
before the gate. The expression for the ideal state $\left| \psi
_{id}\right\rangle $ of the system after the entire gate operation
is straightforward (not shown here to simply our presentation). As
a conservative estimation, we consider $\gamma _{21}^{-1}=\gamma
_{20}^{-1}=\gamma _2^{-1}$ and $\gamma _{10}^{-1}=\gamma _1^{-1}$.
In addition, we choose $g=2\pi \times 50$ KHz [38] and $\kappa
^{-1}=3.0\times 10^{-2}$ s. Our numerical calculation shows that a
high fidelity $\sim 98\%$ can be achieved when the ratio $\Delta
_c/g$ is about 10 (Fig.~5).

For Rydberg atoms chosen here, the $\left| 1\right\rangle
\leftrightarrow \left| 2\right\rangle $ transition frequency is
$\sim 51.1$ GHz [30]. The cavity mode frequency is then $\sim 51.09$
GHz for $\Delta _c/g=10$. For this value of the cavity mode
frequency and the $\kappa ^{-1}=3.1\times 10^{-2}$ s
chosen in our calculation, the required quality factor $Q$ of the cavity is $%
\sim 10^{10}.$ Note that cavities with a high $Q\sim 4.2\times
10^{10}$ have been reported [40]. Our analysis given here shows
that implementing a phase gate with one qubit simultaneously
controlling four target qubits (i.e., a four-target-qubit
controlled phase gate) with atoms is possible within the present
cavity QED technique.

We should mention that the motivation for using
the circular Rydberg states is that they have long energy relaxation times
and have been widely used in quantum information processing [6,30,38,39,41-43].

\begin{center}
\textbf{V. DISCUSSION AND CONCLUSION}
\end{center}

Before conclusion, we should point out that in 2010, Yang, Liu and
Nori proposed a first scheme for implementing a phase gate of one qubit
simultaneously controlling $n$ target qubits based on cavity
QED [44]. In the same year, Yang, Zheng and Nori proposed another scheme for
realizing a multiqubit {\it tunable} phase gate of one qubit simultaneously
controlling $n$ target qubits within cavity QED [45]. As discussed in
[44,45], these two schemes require applying a pulse
to each of qubit systems inside a cavity. The purpose of this work is to
present an alternative approach for implementing the proposed gate. As shown
above, application of a pulse is not required and thus our present scheme
differs from the previous ones in [44,45]. Because of no pulses needed,
the present scheme is much improved when compared with the previous
proposals in [44,45].

We have proposed a way to realize a multiqubit controlled phase gate
with one qubit simultaneously controlling $n$ target qubits using
atoms in cavity QED. As shown above, the gate can be implemented:
(i) by using one cavity through three-step operations only, (ii)
without need of using classical pulses during the gate operations,
(iii) faster when compared with sending atoms through a cavity
or loading atoms into a cavity one by one; and (iv) in an operation
time which scales as $\sqrt{n}$ only. We believe that
this work is of interest because it provides a way for implementing
the proposed multiqubit gate useful in quantum information processing.
Finally, we note that this proposal is quite general, which can be
applied to other physical systems such as various superconducting
qubits coupled to a resonator, NV centers coupled to a microsphere
cavity or quantum dots in cavity QED.

\begin{center}
\textbf{ACKNOWLEDGMENTS}
\end{center}

C.P. Yang was supported in part by the National Natural Science Foundation
of China under Grant No. 11074062, the Zhejiang Natural Science Foundation under
Grant No.~Y6100098, the funds from Hangzhou Normal University, and the Open Fund
from the SKLPS of ECNU. Q.P. Su was supported by the National
Natural Science Foundation of China under Grant No. 11147186.

\end{document}